\documentclass[titlepage,12pt]{article}
\usepackage{amssymb,amsmath,amsfonts}
\usepackage[utf8]{inputenc}
\usepackage{setspace}
\usepackage{epsfig}
\usepackage{graphicx}
\usepackage{cite}
\usepackage{caption}
\usepackage{enumitem}

\makeatletter
\def\@sect#1#2#3#4#5#6[#7]#8{\ifnum #2>\c@secnumdepth
  \def\@svsec{}\else
  \refstepcounter{#1}\edef\@svsec{\csname the#1\endcsname.\hskip0.5em}\fi
  \@tempskipa #5\relax
  \ifdim \@tempskipa>\z@
    \begingroup
      #6\relax
      \@hangfrom{\hskip #3\relax\@svsec}{\interlinepenalty \@M #8\par}%
    \endgroup
    \csname #1mark\endcsname{#7}\addcontentsline
      {toc}{#1}{\ifnum #2>\c@secnumdepth \else
        \protect\numberline{\csname the#1\endcsname}\fi #7}%
  \else
    \def\@svsechd{#6\hskip #3\@svsec #8\csname #1mark\endcsname
      {#7}\addcontentsline{toc}{#1}{\ifnum #2>\c@secnumdepth \else
        \protect\numberline{\csname the#1\endcsname}\fi #7}}%
  \fi \@xsect{#5}}

\newcommand{\hp}{\mathbf{\hat p}}
\newcommand{\hk}{\mathbf{\hat k}}

\newcommand{\ttbar}{t{\bar t}}
\newcommand{\mttbar}{M_{t{\bar t}}}

\newcommand{\Sp}{{\mathbf S}_t}
\newcommand{\Sm}{{\mathbf S_{\bar t}}}
\newcommand{\hlp}{{\mathbf{\hat\ell}_+}}
\newcommand{\hlm}{{\mathbf{\hat\ell}_-}}

\newcommand{\Rmu}{{\rm Re} {\hat\mu}_t}
\newcommand{\Imu}{{\rm Im} {\hat\mu}_t}
\newcommand{\Rd}{{\rm Re} {\hat d}_t}
\newcommand{\Id}{{\rm Im} {\hat d}_t}
\newcommand{\Hmu}{{\hat\mu}_t}
\newcommand{\Hd}{{\hat d}_t}

\begin{document}
\begin{titlepage}
  \begin{flushright}
TTK-13-13
    \end{flushright}
\vspace{0.01cm}
\begin{center}
{\LARGE {\bf Top quark spin correlations and polarization at
    the LHC: \\ standard model predictions and effects of anomalous top chromo
    moments}  \\
\vspace{1.5cm}
\large{\bf Werner Bernreuther\,$^{a}$\footnote{\tt breuther@physik.rwth-aachen.de} and 
 Zong-Guo Si\,$^{b}$}\footnote{\tt zgsi@sdu.edu.cn}
\par\vspace{1cm}
$^a$Institut f\"ur Theoretische Physik, RWTH Aachen University, 52056 Aachen, Germany\\
$^b$Department of Physics, Shandong University, Jinan, Shandong
250100, China
\par\vspace{1cm}
{\bf Abstract}\\
\parbox[t]{\textwidth}
{\small{ A number of top-spin observables are computed within the
    Standard Model (SM),
    at next-to-leading order in the strong and weak gauge couplings for
    hadronic top-quark anti-quark ($\ttbar$) production and decay at
    the  LHC for center-of-mass energies 7 and 8 TeV. For dileptonic
    final states we consider the azimuthal angle correlation, the
    helicity correlation, and the opening angle distribution; for
    lepton plus jets final states we determine distributions and
    asymmetries that trace a longitudinal and transverse polarization,
    respectively, of the $t$ and $\bar t$ samples. 
   In addition, we investigate the effects of a non-zero
   chromo-magnetic and chromo-electric dipole moment of the top quark
   on these and other   top-spin observables and associated
   asymmetries. These observables allow to disentangle the 
   contributions from the real and imaginary parts of these moments. 
}}
}
\end{center}
\vspace*{0.7cm}

PACS number(s): 12.38.Bx, 13.88.+e, 14.65.Ha
\end{titlepage}
%
%
\setcounter{footnote}{0}
\renewcommand{\thefootnote}{\arabic{footnote}}
\setcounter{page}{1}

\section{Introduction} 
\label{introduction}

Last year the ATLAS collaboration \cite{ATLAS:2012ao} at the LHC measured the correlation
of $t$ and $\bar t$ spins in $\ttbar$ production at the LHC. 
 The hypothesis of zero spin correlation was  excluded at 5.1 standard
 deviations. (See also the CMS analysis \cite{CMS12spin}.) Previously,
 the D$\emptyset$ collaboration \cite{Abazov:2011gi} found 
 evidence for   $\ttbar$ spin correlations in  
events with a significance of more than 3 standard deviations.
These experimental results at the LHC and at the Tevatron are in
agreement, within  uncertainties, with corresponding
standard model (SM) predictions, and therefore provide another experimental proof that the
 top quark behaves like a bare quark that does not hadronize.

In view of these findings, top spin observables are rather unique tools (as
compared to corresponding observables for lighter quarks) for the
detailed exploration of, in
particular,   top-quark pair  production (and decay) dynamics, because 
 (future) measurements of angular correlations/distributions induced
 by $\ttbar$ spin correlations or $t$, $\bar t$ polarization can be
 confronted with reliable perturbative predictions within the SM
 versus predictions made with new-physics (NP) models.

As to the modelling of new physics effects one may either consider a
specific NP model, e.g. the minimal supersymmetric extension of the SM,
or use a rather model-independent approach to parameterize possible NP
effects in  top-quark production and decay. Here we shall use the
second approach.

We consider $\ttbar$ production at the LHC and subsequent decays into
 dileptonic and lepton plus jets final states. The aim of this paper is
 twofold. On the one-hand we extend our previous SM predictions
 of a number of  $\ttbar$ spin-correlation effects 
 at next-to-leading
 order in the strong and weak gauge couplings \cite{Bernreuther:2001rq,Bernreuther:2004jv,Bernreuther:2010ny}, and of
  SM-induced longitudinal \cite{Bernreuther:2006vg,Bernreuther:2008md}
  and transverse \cite{Bernreuther:1995cx} $t$ and $\bar t$ polarization
  to $pp$ collisions at the LHC at center-of-mass energies of 7 and 8 TeV. In addition,
  we analyze the effects of a chromo-magnetic and
   chromo-electric dipole moment of the top quark on  $\ttbar$
   spin correlations and on $t$ and $\bar t$ polarization. 

 Assuming that new physics effects in hadronic  $\ttbar$ 
 production are induced by new heavy particle exchanges (characterized
 by a mass scale
 $M$) one may
 construct a local effective Lagrangian ${\cal L}_{eff}$ that respects the SM gauge symmetries
 and describes possible new physics interaction structures for
 energies $\lesssim M$. Recent analyses include
 \cite{AguilarSaavedra:2008zc,AguilarSaavedra:2009mx,Zhang:2010dr,AguilarSaavedra:2010zi,Grzadkowski:2010es,Degrande:2010kt}.
 Here we confine ourselves to interactions of mass dimension 5 after 
  spontaneous electroweak symmetry breaking. Then, as is well-known,
  the new-physics part of ${\cal L}_{eff}$ is given in terms of
 chromo dipole couplings of the top quark to the gluon(s):
 \begin{equation}\label{Leff} 
{\cal L}_{eff} = {\cal L}_{SM} - \frac{{\tilde\mu}_t}{2} {\bar t}\sigma^{\mu\nu}T^a t G^a_{\mu\nu}
  - \frac{{\tilde d}_t}{2} {\bar t}i\sigma^{\mu\nu}\gamma_5
  T^a t G^a_{\mu\nu} \, ,
\end{equation}
  where ${\tilde\mu}_t$  and  ${\tilde d}_t$ 
are the chromo-magnetic  (CMDM)  and  chromo-electric (CEDM) dipole moment 
  of the top quark, respectively, $G^a_{\mu\nu}$ denotes the gluon
  field strength tensor, and  $T^a$ the generators of $SU(3)$ color. 
In particular, a sizeable non-zero CEDM would signal a  new type of
CP-violating interaction beyond the Kobayashi-Maskawa CP phase.

It is customary to define dimensionless chromo moments $\Hmu, \Hd$ by
\begin{equation}\label{Defcmud}
{\tilde\mu}_t =\frac{g_s}{m_t} \Hmu \, ,
  \qquad {\tilde d}_t =\frac{g_s}{m_t} \Hd \, ,
\end{equation}
where $m_t$ denotes the top-quark mass and $g_s$ is the QCD coupling.

There exists an extensive literature on the phenomenology of anomalous  top-quark
chromo  moments in hadronic $\ttbar$ production \cite{Atwood:1992vj,Brandenburg:1992be,Atwood:1994vm,Haberl:1995ek,Cheung:1995nt,Cheung:1996kc,Grzadkowski:1997yi,Yang:1997iv,Hikasa:1998wx,Zhou:1998wz,Atwood:2000tu,Lillie:2007hd,Gupta:2009eq,Gupta:2009wu,Hioki:2009hm,HIOKI:2011xx,Choudhury:2009wd,Bach:2012fb,Gabrielli:2012pk}.
The topic has been revisited  recently
\cite{Hioki:2012vn,Englert:2012by,Biswal:2012dr,Baumgart:2012ay,Hesari:2012au,Baumgart:2013yra}
 in view of the large samples
 of $\ttbar$ events that have been recorded so far at the LHC. In
 fact, the analysis of these data samples  yields  useful direct
 information\footnote{For indirect upper bounds on 
   $|\Hd|$, $|\Hmu|$, cf. \cite{Martinez:1996cy,Kamenik:2011dk}.}
 on $\Hmu$ and $\Hd.$ For instance, a
 comparison of $\sigma_{\ttbar}^{exp}$ with the SM prediction, made in
       \cite{Hioki:2012vn},     yields
  a region in the set  of couplings $\Hmu, \Hd$ that is still allowed.
  This allowed region is given roughly by $|\Hmu|\lesssim 0.03$,
  $|\Hd|\lesssim 0.1$. We will use this result in our analysis below,
  i.e., we use that the moduli of the dimensionless chromo moments
 of the top quark, if non-zero at all, are markedly smaller than one.
 
Because the chromo moments $\Hmu$ and $\Hd$ arise from respective form
factors in the limit of large $M$, we slightly extend the framework of
 \eqref{Leff} in our analysis below and take into account that these
 form factors may have absorptive, i.e., imaginary parts if the
 4-momentum transfer $q^2$ in a gluon-top vertex is timelike, in
 particular if $q^2>4 m_t^2$. 
 Therefore, we use in the following the parameterization 
\begin{equation}\label{Defgenp}
\Hmu  = \Rmu + i \Imu, \qquad 
 \Hd = \Rd + i \Id   \, ,
\end{equation}
and allow for imaginary parts if the 4-momentum transfer
 $q^2>4 m_t^2$ in the respective gluon-top vertex.
 We emphasize that $\Hmu, \Hd$ parameterize by definition only
 new physics contributions to $gtt$ and $ggtt$ vertices.
 The dependence on $q^2$ of $\Hmu, \Hd$ depends on the specific new
 physics model.
We assume that $\Hmu, \Hd$ are constants.
As we consider below only normalized top-spin observables, this
assumption does not spoil perturbative unitarity. 

In the following section we consider $\ttbar$ production and
decay into dilepton and lepton plus jets final states at the LHC (7
and 8 TeV). We compute the 
 contributions  of $\Hmu, \Hd$ to the respective matrix elements
 that are linear in the chromo moments. This linear approximation is
  justified by the  upper bounds cited above. We show in
  Sect.~\ref{sec:obsR} that 
 the contributions of $\Rmu, \Rd, \Imu,$ and $\Id$ can be disentangled
 with  appropriate $\ttbar$ spin correlation and top polarization
 observables. In addition, we compute also  distributions, expectation
 values, and asymmetries of these spin observables  at next-to-leading
 order in the strong and weak gauge couplings. In particular, we
 recompute the SM-induced transverse polarization of the $t$ and $\bar
 t$ quarks, an effect that is worth  to be investigated in its own right.

 \section{Set-up of the computation}

We consider $\ttbar$ production  at the LHC and subsequent decay into  dileptonic
final states, 
\begin{equation}\label{ppdilep}
p p \to t + {\bar t} + X\to \ell^ + \ell'^-  + \ \text{jets}  + E_T^{\rm miss}\, ,
\end{equation}
 and into  lepton plus jets final states,
\begin{eqnarray}
p p & \to t + {\bar t} + X & \to \ell^+ +  \ \text{jets} + E_T^{\rm miss}\, , \label{ppljt}\\
p p & \to t + {\bar t} + X& \to \ell^- + \ \text{jets} + E_T^{\rm miss} \, .\label{ppljtbar}
\end{eqnarray} 
where $\ell,\ell'= e,\mu,\tau.$ We use the narrow width approximation for the top quark.
 Within the SM we consider $gg, q{\bar q}$, $gq$, and $g\bar q$ initiated $\ttbar$ production
 at next-to-leading order in the strong and weak couplings, taking the $t$ and $\bar t$
 spin degrees of freedom fully into account. On-shell top-quark decay is incorporated 
 at next-to-leading order in the strong coupling in a consistent way. Our computational procedure
  is described in detail in \cite{Bernreuther:2010ny}. We refer to this perturbative calculation of the 
respective parton matrix elements by the acronym NLOW.

 As justified above,  we take top-quark chromo moments $\Hmu, \Hd$  into account only in the linear approximation in
 the following. That is, we consider the interference of the
 leading-order (LO)  QCD amplitudes, i.e., the LO amplitudes 
 for $gg, q{\bar q} \to t {\bar t}$ with the corresponding amplitudes that contain a
  chromo-moment $\Hmu$ or $\Hd$. In the case of $ q{\bar q}$ initial states, this
  interference term $\delta{\cal M}^{NP}_{q{\bar q}}$ results from the interference of
  the SM and the corresponding NP s-channel diagram. As $s>4m_t^2$,
  the chromo form factors may have imaginary parts that 
  we take  into account.  The LO amplitude of $gg \to t {\bar t}$ involves $s$-, $t$-, and  $u$-channel
  diagrams. Only the chromo form factors that are associated with the $s$-channel diagrams   may have an imaginary part.
 In this way 
 the interference terms $\delta{\cal M}^{NP}_i$ ($i=gg,q{\bar q}$) are obtained
 that depend linearly on $\Rmu, \Rd, \Imu,$ and $\Id$ and the $t$, $\bar t$ spins. From these terms we extract the NP contributions $\delta R_i^{NP}$
 to the corresponding SM production density matrices computed at NLOW \cite{Bernreuther:2010ny}.
 The production density matrices $\delta R_i^{NP}$ were computed,
 quite some time ago, for $\Rd\neq 0$ by \cite{Brandenburg:1992be} and
  for  $\Rmu, \Rd\neq 0$ by \cite{Haberl:1995ek}. Our analytic  results for
  $\delta R_i^{NP}$ agree with those of \cite{Brandenburg:1992be,Haberl:1995ek}.

 The NP contributions to the parton matrix elements that describe $\ttbar$ production and decay into dileptonic and lepton plus
  jets final states involve also the SM top-quark decay 
    density matrices for $t\to W b \to \ell\nu b, q {\bar q}' b$ to LO.     
  At this point the following remark is in order. The top-decay vertex $t\to W b$ may also be affected by new physics
 interactions that can also be parameterized by anomalous couplings.  We use in the following  as  spin-analyzers
  of the (anti)top quark only the charged lepton $\ell^\pm$ from $W^\pm$ decay, and we consider below only lepton angular correlations and
  distributions that are inclusive in the lepton energies. It is known
  (see e.g. \cite{Rindani:2000jg,Grzadkowski:1999iq,Grzadkowski:2002gt,Godbole:2006tq})
 that these observables are not affected by anomalous couplings
 from top-quark decay if these couplings are small, i.e., if a linear approximation is justified.
 This is indeed the case in view of the present upper  bounds  on the
 moduli of these couplings that can be inferred from the measured
 $W$-boson helicity fractions in top-quark decay (cf., e.g.,
 \cite{Yazgan:2013jma}). In other words, the observables 
  that we analyze in the next section are affected only by possible new physics
   contributions to $\ttbar$ production that we parameterize  by complex chromo-moments
  $\Hmu, \Hd$.

 For the computations below we use the following input parameters:
 $m_t=  173.1 \ {\rm GeV},  \Gamma_t  =1.3  \ {\rm GeV},$ $m_W=  80.4 \ {\rm GeV}, 
 \Gamma_W = 2.09 \ {\rm GeV}, \alpha_s(\mu=m_Z)= 0.112,$
 and we use the CTEQ6.6M parton distribution functions \cite{Nadolsky:2008zw}.

The observables of Sect.~\ref{sec:obsR}  involve the following
inertial frames: i) the laboratory frame which is
defined by using one of the proton beams as the $z$ axis and choosing
the orthogonal $x$ and $y$ axes such that a right-handed coordinate
system results. ii) The $\ttbar$ zero-momentum frame (ZMF) is
obtained  by a rotation-free boost from the laboratory frame.
iii) The $t$ and $\bar t$ rest frames are obtained by respective
rotation-free boosts from the $\ttbar$ ZMF.

 \section{Observables and results}
\label{sec:obsR}

We analyze a set of top-spin observables that allow to
 disentangle, in the linear approximation for the top chromo
 moments\footnote{An analysis of the transformation
  properties of the contributions of $\Rmu, \Rd, \Imu,$ and $\Id$
  to the matrix elements with respect to parity, charge conjugation,
 and naive time `reversal' $T_N$ (reversal of spins and 3-momenta)
 shows that these terms can be disentangled with the top-spin
 observables used in this section. For a general analysis, see \cite{Bernreuther:1993hq}.},
 the contributions from $\Rmu, \Rd, \Imu,$ and $\Id$ to the
 matrix elements of  \eqref{ppdilep} - \eqref{ppljtbar}.
 Because the charged lepton form top decay is the best top-spin analyzer,
 we consider here 
  only lepton angular correlations and distributions.

  For the dileptonic final states  \eqref{ppdilep} we consider the following
  observables that involve the charged leptons
 $\ell^+, \ell'^-$: i) the azimuthal angle correlation,
  the helicity correlation and the opening angle distribution. 
    If the
  top chromo-moments are non-zero, they receive a contribution
  proportional to $\Rmu$.
  ii) Two CP-odd triple correlations that are
     sensitive to $\Rd$. 
      
The lepton plus jets final states \eqref{ppljt}, \eqref{ppljtbar} are
the most suitable channels for checking whether or not the $t$ and
$\bar t$ of the hadronically produced $\ttbar$ sample have a sizeable
polarization. A non-zero $\Id$  induces  a longitudinal $t$ and
$\bar t$ polarization which the SM predicts to be very small.
 A non-zero $\Imu$ contributes to the transverse polarization of $t$
 and $\bar t$ which in the SM is generated predominantly by QCD
 absorptive parts of the scattering amplitudes.

As it has become customary in experimental analyses to present
results also by unfolding data for comparison with predictions made at the
level of final state partonic jets and/or leptons, we do not, in the
 following, apply acceptance cuts to the final states in
 \eqref{ppdilep} and \eqref{ppljt}, \eqref{ppljtbar}.

 \subsection{Observables for tracing $\Rmu$}
\label{suse:Rmu}

 We consider the dileptonic final states \eqref{ppdilep} and analyze
 first  the normalized distribution of 
 the difference of the azimuthal angles of the charged leptons in the
 laboratory frame \cite{Arens:1992fg,Mahlon:2010gw,Bernreuther:2010ny}, $\Delta\phi=\phi_{\ell^+} - \phi_{\ell^-}$. 
\begin{equation}\label{fundeq}
\sigma^{-1}\frac{d\sigma}{d\Delta\phi} = 
\sigma^{-1}\frac{d\sigma_{SM}}{d\Delta\phi} + 
\sigma^{-1}\frac{d\sigma_{NP}}{d\Delta\phi} \, . 
\end{equation}
As emphasized above, we take into account in 
    $d\sigma_{NP}$ only the contributions linear in the chromo moments. As
  we compute  $d\sigma_{SM}$ to NLO in the SM couplings, we have  for
  the integrated cross section of  \eqref{ppdilep}:
\begin{equation} \label{xsecSNP}
\sigma =\sigma_{LO} + \sigma_{NLOW} + \sigma_{NP} \, ,
 \end{equation}
where, in the linear approximation, the term  $\sigma_{NP}$ (which is
 in general not positive) receives only a contribution from
   $\Rmu$, i.e.,  $\sigma_{NP}= {\cal O}(\alpha_s^2 \Rmu)$.

For the calculation of a ratio like \eqref{fundeq}  to
 NLO in the SM couplings 
 one has two options: expanding or not expanding the denominator. We use here as default procedure the first
  option, which is common practice in higher-order perturbative
  calculations.
   We then get for the first term on the right-hand side of \eqref{fundeq}:
\begin{eqnarray}
\frac{1}{\sigma}\frac{d\sigma_{SM}}{d\Delta\phi} & = &
 \frac{1}{\sigma_{LO}} \frac{d\sigma_{LO+NLOW}}{d\Delta\phi}
   -  \frac{\sigma_{NLOW}}{\sigma_{LO}^2}\frac{
     d\sigma_{LO}}{d\Delta\phi} 
 -  \frac{\sigma_{NP}}{\sigma_{LO}^2}\frac{
     d\sigma_{LO}}{d\Delta\phi} 
\,  + \,  {\cal O}(\alpha_s^2) \, , \nonumber \\
 & \equiv & \left(\frac{1}{\sigma}\frac{d\sigma}{d\Delta\phi}\right)_{SM} -  \frac{\sigma_{NP}}{\sigma_{LO}^2}\frac{
     d\sigma_{LO}}{d\Delta\phi} \,  + \,  {\cal O}(\alpha_s^2) \, .
\label{SMpexp}
\end{eqnarray}
Here and below, the label $SM$ refers to the
  LO and NLOW contributions, i.e.,
 the first two terms in the first line of  \eqref{SMpexp}.
 The contribution in \eqref{SMpexp} proportional to  $\sigma_{NP}$ will
be added to  the
 expanded second term on the right-hand side of \eqref{fundeq}. The
 total NP  contribution 
  is then given 
 to order  $\Rmu$  by 
\begin{equation}\label{npexp}
\left(\frac{d{\sigma}}{\sigma d\Delta\phi}\right)_{NP} \Rmu \equiv
\frac{1}{\sigma_{LO}}\left(\frac{d\sigma_{NP}}{d\Delta\phi}
 - \frac{\sigma_{NP}}{\sigma_{LO}}\frac{d\sigma_{LO}}{d\Delta\phi}
\right)  \, .
\end{equation}
The expanded form \eqref{npexp} is convenient as it is proportional
 to $\Rmu$.
In summary, we compute the  right-hand side of \eqref{fundeq} by
 computing the sum of $(\sigma^{-1}d\sigma/d\Delta\phi)_{SM}$ (cf. \eqref{SMpexp})  and \eqref{npexp}.
 Notice that the integral $\int  d\Delta\phi$ over this sum is one,
  as it should be. 

\begin{figure}
\centering
\includegraphics[width=11cm,height=11cm]{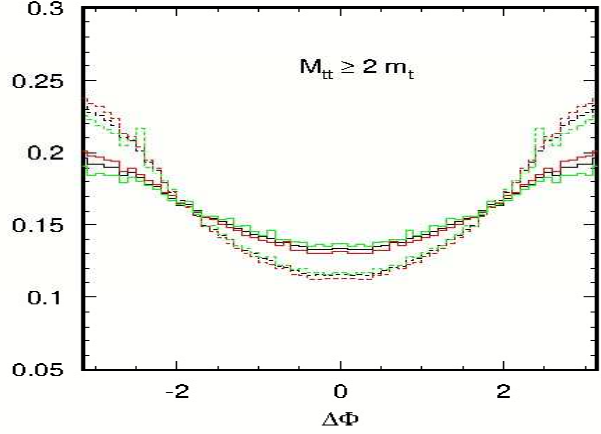}
\includegraphics[width=11cm,height=11cm]{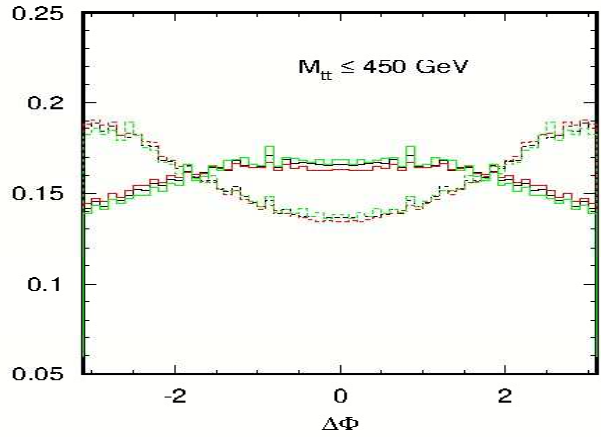}
\caption{
\label{fig:dphiSM7}
 SM prediction  $(\sigma^{-1}d\sigma/d\Delta\phi)_{SM}$ defined in
 \eqref{SMpexp}, \eqref{DphiSMNP}
                    at NLOW for the normalized dilepton  $\Delta\phi$ distribution
 at the LHC (7 TeV).
  Dashed = uncorrelated, solid = correlated. The chosen
 scales are $\mu=m_t$(black), $2m_t$ (red), and $m_t/2$
 (green). (Color code in online version only.)
 Upper plot: distribution without cut on $\mttbar$.
 Lower plot: distribution for events
  with $\mttbar\leq 450$ GeV.
 }
\end{figure}

In Figs.~\ref{fig:dphiSM7} the SM contribution to NLOW of the $\Delta\phi$ distribution 
  is shown\footnote{SM predictions for the distribution of
    $\Delta\phi$ and of the other observables below for 7 and 8 TeV  can be obtained
    from the authors upon request.}    for $\sqrt{S_{had}}=7$ TeV for no cut
  on the $\ttbar$ invariant mass $\mttbar$  and for events
  with $\mttbar\leq 450$ GeV.
  (This cut  was chosen in the experimental
  analysis \cite{CMS12spin}.)  The distributions are symmetric with respect
 to $\Delta\phi=0$.
 For reference, the  $\Delta\phi$ distribution 
  is shown in these figures also for the case when the $\ttbar$ spin correlations
  are switched of. These distributions were used in  \cite{ATLAS:2012ao,CMS12spin} for testing
   the hypothesis ``SM, fully correlated'' versus ``SM, uncorrelated''
   which led to the exclusion of the second hypothesis with 5.1 s.d. \cite{ATLAS:2012ao}.
  In the following, we do no longer consider this option, i.e., we
   will always consider  fully spin-correlated $\ttbar$ events, both for
   SM and NP predictions.

\begin{figure}
\centering
 \includegraphics[width=12cm]{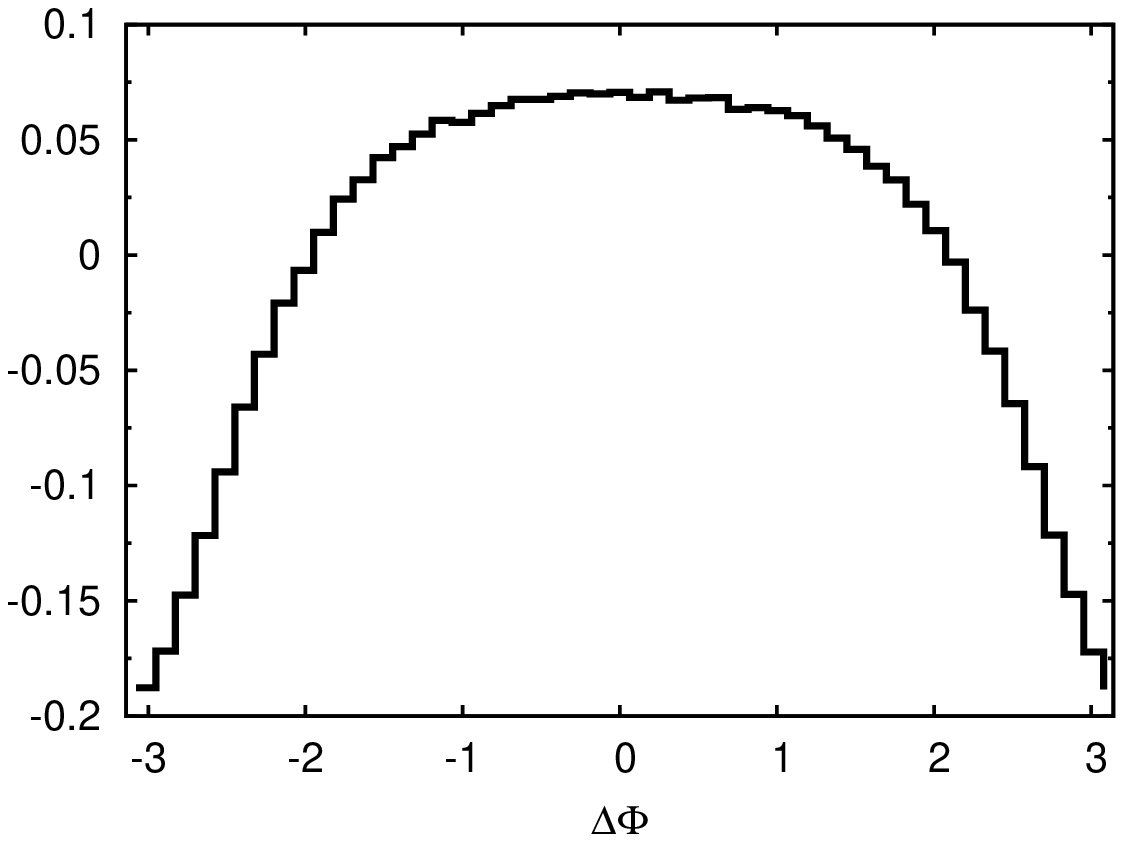}
 \includegraphics[width=12cm]{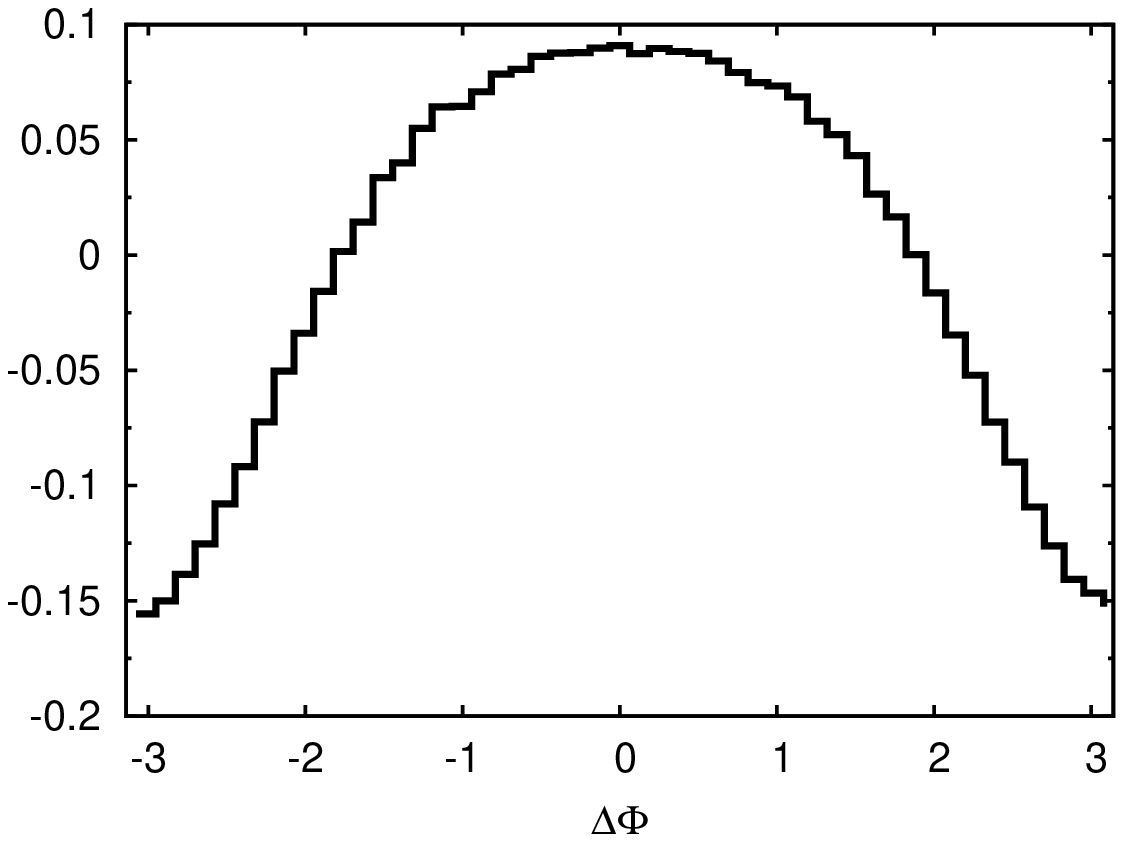}
\caption{
\label{fig:dphiChr7NP}
 The contribution  $(\sigma^{-1}d\sigma/d\Delta\phi)_{NP}$ defined in
\eqref{npexp}   to the  $\Delta\phi$ distribution
 at the
LHC (7 TeV). Upper plot: no cut on $\mttbar$. Lower plot: 
   events with    $\mttbar< 450$  GeV. The thickness of the
 histogram bars reflects the effects of scale variations
 $\mu=m_t/2,m_t, 2m_t$.}
\end{figure}

 For $\Rmu\neq 0$ and $|\Rmu|\ll 1$ we get 
\begin{equation} \label{DphiSMNP}
\frac{1}{\sigma}\frac{d\sigma}{d\Delta\phi} =\left(\frac{1}{\sigma}\frac{d\sigma}{d\Delta\phi}\right)_{SM} +
\left(\frac{1}{\sigma}\frac{d\sigma}{d\Delta\phi}\right)_{NP} \Rmu  \, .
\end{equation}
The contribution $(\sigma^{-1}d\sigma/d\Delta\phi)_{NP}$ is shown in
Figs.~\ref{fig:dphiChr7NP} for $\sqrt{S_{had}}=7$ TeV
  in the range $-\pi< \Delta\phi\leq \pi$ for  no cut on $\mttbar$ 
  and for events with  $\mttbar\leq 450$ GeV. (For $\mttbar>450$ GeV
  this contribution has the same shape as those of
  Figs.~\ref{fig:dphiChr7NP} and is therefore not shown here.) 
  As expected, these contributions are  also symmetric
   with respect to  $\Delta\phi=0$. (Thus, the SM and NP numbers must be doubled if the range
   $0\leq \Delta\phi\leq \pi$ is considered.) Furthermore, by the above assumption,
   $|\Rmu|$ must be sufficiently small such that the sum of the two
   terms on the right-hand side of 
 \eqref{DphiSMNP} is positive. Adding these contributions to the respective correlated SM
  contributions shown in  Figs.~\ref{fig:dphiSM7} we see that a cut on $\mttbar$ is of no advantage.
  Thus, when the   $\Delta\phi$ distribution is used
 to probe for a non-zero $\Rmu$ one should use the full sample of dileptonic $\ttbar$ events.

One may also probe for a non-zero  CMDM $\Rmu$ with the dileptonic angular
  correlation in the helicity basis. If no acceptance cuts are applied, one has
 the well-known a priori form of the double angular distribution
\begin{equation}\label{doublehel}
 \frac{1}{\sigma} \frac{d\sigma}{d\cos\theta_1d\cos\theta_2}= \frac{1}{4}\left(1+ B_1\cos\theta_1
   + B_2\cos\theta_2 - C \cos\theta_1  \cos\theta_2        \right) \, ,
\end{equation}
where, in the   helicity basis, 
$\hlp$ ($\hlm$)  is the $\ell^+$ ($\ell^-$) direction of flight in the
$t$  ($\bar t$) rest frame, 
${\bf\hat k}_t$ and ${\bf\hat k}_{\bar t}=-{\bf\hat k}_t$ are the  $t$
and $\bar t$ directions of flight in the $\ttbar$ ZMF, respectively,
  and $\theta_1 =\angle(\hlp,{\bf\hat k}_t)$, $\theta_2 =\angle(\hlm,{\bf\hat k}_{\bar t})$.
 For the experimental analysis it is more convenient to use the one-dimensional
  distributions \cite{Bernreuther:2004jv,Bernreuther:2010ny} of the
  product of the cosines  ${\cal O}_h \equiv \cos\theta_1\cos\theta_2$, rather than analyzing \eqref{doublehel}.
In the linear approximation for the chromo moments, we get in analogy to   \eqref{DphiSMNP}:
\begin{equation} \label{DOhelSMNP}
\frac{1}{\sigma}\frac{d\sigma}{d{\cal O}_h}
=\left(\frac{1}{\sigma}\frac{d\sigma}{d{\cal O}_h}\right)_{SM} +
\left(\frac{1}{\sigma}\frac{d\sigma}{d{\cal O}_h}\right)_{NP} \Rmu \, .
\end{equation}

\begin{figure}
\centering
\includegraphics[width=11cm,height=9cm]{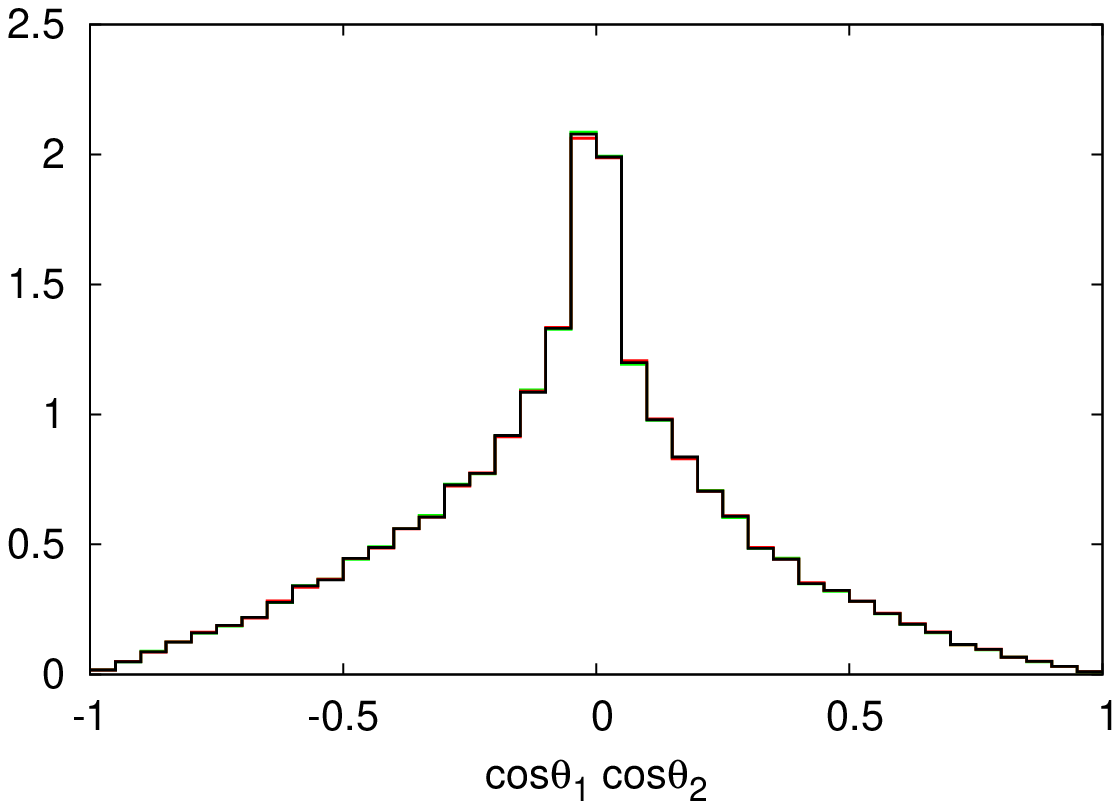}
\includegraphics[width=11cm,height=9cm]{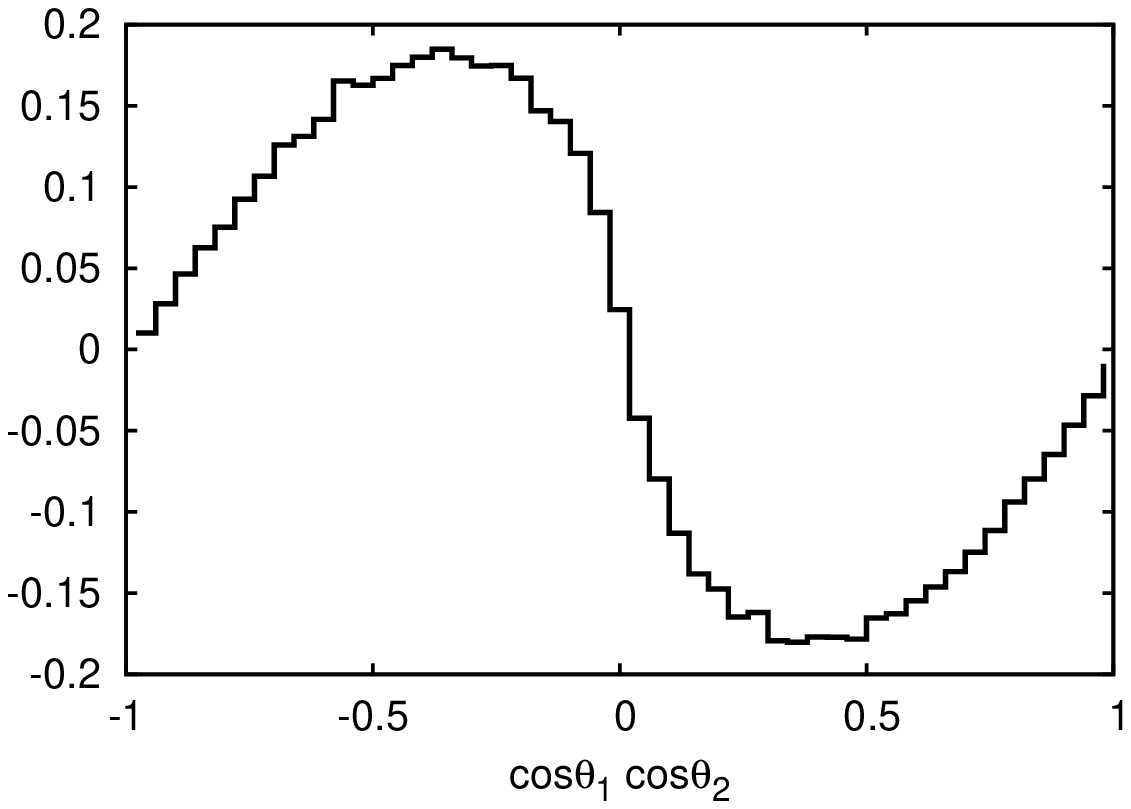}
\caption{
\label{fig:HelSMNP7}
The normalized distribution of the product $\cos\theta_1\cos\theta_2$
of the lepton helicity angles, defined in
       \eqref{DOhelSMNP}     for the LHC (7 TeV).
 Upper plot:  SM prediction at NLOW, $({\sigma}^{-1}{d\sigma}/{dO_h})_{SM}$.
The chosen
 scales are $\mu=m_t$(black), $2m_t$ (red), and $m_t/2$
 (green). (Color code in online version only.)
 Lower plot: NP contribution $({\sigma}^{-1}{d\sigma}/{dO_h})_{NP}$.
 The thickness of the
 histogram bars reflects the effects of scale variations
 $\mu=m_t/2,m_t, 2m_t$. }
\end{figure}

In Figs.~\ref{fig:HelSMNP7} the SM contribution to NLOW
 and the NP contribution  to this  distribution 
  are  shown for $\sqrt{S_{had}}=7$ TeV, for no cut
 on  $\ttbar$. Again, we emphasize that   \eqref{DOhelSMNP}
  applies to values of $\Rmu$ such that the distribution is positive.
We find that applying an upper or lower cut on $\mttbar$,
 for instance,  $\mttbar\leq 450$ GeV or $\mttbar> 450$ GeV,
 does not increase the sensitivity of this distribution to  
    $\Rmu$ significantly.

The correlation coefficient $C$ in \eqref{DOhelSMNP} is given 
   in the helicity basis by
\begin{equation}\label{ChelSMNP}
 C_{hel} = -9 \langle  \cos\theta_1  \cos\theta_2    \rangle =
C^{SM}_{hel} + C^{NP}_{hel} \Rmu  \, .
\end{equation}
Our results for $C^{SM}_{hel}$ and $C^{NP}_{hel}$ are given
 in Table~\ref{table:chel}. The label `NLOW, expanded' refers to the
 computation of $C^{SM}_{hel}$ in analogy to  \eqref{SMpexp}.
 For reference purposes, with regard to experimental analyses, we have determined $C^{SM}_{hel}$ also
 by not expanding the denominator, i.e., by computing 
$\langle{\cal O}_h\rangle =\int d\sigma_{LO+NLOW}{\cal O}_h/(\sigma_{LO} + \sigma_{NLOW}).$

One may also define an asymmetry which, in the absence of acceptance
cuts, is determined by   $C_{hel}$:
\begin{equation}\label{Asyhel}
A_{h} =
\frac{N_{\ell\ell}({\cal O}_h >0)-N_{\ell\ell}({\cal O}_h <0)}{N_{\ell\ell}({\cal O}_h>0)+N_{\ell\ell}({\cal O}_h<0)}
 = - \frac{C_{hel}}{4} \, .
\end{equation}

\begin{table}[htb]
\caption{The contributions to the spin correlation coefficient $C_{hel}$,
  defined in \eqref{ChelSMNP},  for dileptonic events at
  the LHC (7 and 8 TeV)  and the scale choice  $\mu=m_t$. The
  uncertainties in parentheses result from scale choices $\mu=m_t/2, 2m_t.$ }
\label{table:chel}
\begin{center}
\begin{tabular}{lccc}
\hline \hline 
7 TeV  & $\mttbar\geq 2 m_t$ & $\mttbar \leq 450$ GeV    & $\mttbar > 450$ GeV \\
\hline 
$C^{SM}_{hel}$ (NLOW)  expanded & $0.310(6)$ & $0.422(2)$ & $0.203(8)$ \\
 $C^{SM}_{hel}$ (NLOW)  unexpanded &$0.295(20)$& $0.417(10)$ & $0.185(22)$
\\
$C^{NP}_{hel}$ &  $0.980(10)$  & $0.972(14) $ &   $0.906(15) $  \\
\hline 
8 TeV  & & & \\
\hline 
$C^{SM}_{hel}$ (NLOW)  expanded & $0.318(5)$ & $0.442(2)$ & $0.205(8)$
\\
 $C^{SM}_{hel}$ (NLOW)  unexpanded & $0.304(14)$ & $0.435(5)$ &
 $0.190(20)$ \\
$C^{NP}_{hel}$ & $0.964(10)$  & $0.949(13)$ & $0.888(10)$  \\
\hline \hline
\end{tabular}
\end{center}
\end{table}
 
Next we consider the opening angle distribution for dileptonic final
states \cite{Bernreuther:1998qv,Bernreuther:2004jv,Bernreuther:2010ny}:
\begin{equation}\label{eq:open-compl}
\frac{1}{\sigma} \frac{d\sigma}{d\cos\varphi}= \frac{1}{2}\left(1- D
  \cos\varphi \right) \, ,
\end{equation}
where $\varphi =\angle(\hlp,\hlm)$ and, as above, 
$\hlp$ $(\hlm)$ is the $\ell^+$ $(\ell^-)$  is  direction of flight in
$t$ $({\bar t})$ rest frame. If  no acceptance cuts are applied then
\begin{equation}\label{openD}
 D = -3\langle \cos\varphi \rangle =  D_{SM} + D_{NP} \Rmu  \qquad
\text{for}\; |\Rmu|\ll 1. 
\end{equation}
An associated asymmetry is
\begin{equation}\label{openAsy}
A_{\varphi} =
\frac{N_{\ell\ell}(\cos\varphi>0)-N_{\ell\ell}(\cos\varphi<0)}{N_{\ell\ell}(\cos\varphi>0)+N_{\ell\ell}(\cos\varphi<0)}
= - \frac{D}{2} \, .
\end{equation}
Our results for the  SM (at NLOW) and NP contributions to the
correlation coefficient $D$ at the  LHC (7 and 8 TeV) are given in
Table~\ref{table:dopen}. As in the case of the SM predictions for
 the  helicity correlation $C^{SM}_{hel}$, we give
 the predictions for $D_{SM}$ at NLOW both in the expanded and in the
 unexpanded form.

\begin{table}[htb]
\caption{The SM and NP contributions to the spin correlation coefficient $D$, defined in
  \eqref{eq:open-compl} and \eqref{openD},  for dileptonic events at
  the LHC (7 and 8 TeV) and the scale choice  $\mu=m_t$. The
  uncertainties in parentheses result from scale choices $\mu=m_t/2, 2m_t.$
 }
\label{table:dopen}
\begin{center}
\begin{tabular}{lccc}
\hline \hline 
7 TeV  & $\mttbar\geq 2m_t$ & $\mttbar \leq 450$ GeV    & $\mttbar > 450$ GeV \\
\hline 
$D_{SM}$ (NLOW)  expanded & $-0.223(4)$ & $-0.332(2)$ & $-0.120(6)$ \\
$D_{SM}$ (NLOW)  unexpanded & $-0.212(12)$& $-0.323(7)$ & $-0.110(15)$
\\
$D_{NP}$  & $-1.675(20)$ &   $-1.670(17)$ & $-1.613(22)$ \\
\hline 
8 TeV & &  & \\ \hline
$D_{SM}$ (NLOW)  expanded & $-0.228(5)$ & $-0.336(2)$ & $-0.130(5)$ \\
$D_{SM}$ (NLOW)  unexpanded & $-0.217(11)$ & $-0.330(6)$ & $-0.120(14)$
\\
$D_{NP}$  & $-1.712(19)$  & $-1.696(14)$ & $-1.653(20)$ \\
\hline \hline 
\end{tabular}
\end{center}
\end{table}

The distributions  \eqref{DphiSMNP},  \eqref{DOhelSMNP}, and
\eqref{eq:open-compl}
 may be used for 1-parameter fits to the respective unfolded
 experimental distributions that ATLAS or CMS may obtain from
 the existing 7 and 8 TeV dileptonic data samples.
  In view of the results given in Tables~\ref{table:chel} and~\ref{table:dopen}
  one may worry that a significant source of theoretical uncertainty
  is  how  the higher-order SM
  contributions are taken into account. As mentioned above, we
  advocate to use the expanded form of the normalized distributions
  for fits to the unfolded data, as this is in the spirit of
  perturbation theory. 
With which uncertainty may $\Rmu$ be measured? The highest sensitivity
 to this parameter will certainly result from fits to these
 distributions,  which is an experimental task. 
  For instance, CMS has reconstructed $\sim 9000$ dilepton events
  $(\ell=e,\mu)$ from  the 7 TeV ($5$ fb$^{-1}$) data \cite{CMS:2012owa}. Assuming that
  the same selection efficiency applies  to the 8 TeV data, one
  expects  $\sim 5\times 10^4$ reconstructed dilepton events from the
  8  TeV ($20$ fb$^{-1}$) data. These numbers suggest that a
  statistical error $\delta\Rmu$ below the percent level is feasible;
  the limiting factor will be the systematic experimental and theoretical
  uncertainties. A crude estimate may be done with the asymmetries
  introduced above. They should be rather robust from the experimental
  point of view, but contain, of course, less information than
  the  underlying distributions. Using for instance the asymmetry \eqref{openAsy},
   which has a rather
  large lever arm to $\Rmu$, and assuming that $A_{\varphi}$ may be
  measured at 8 TeV with a combined statistical and systematic  uncertainty $\delta A_{\varphi}
  =0.03$, then $\Rmu$ may be extracted with an uncertainty 
  $\delta \Rmu \simeq 0.04$. Estimates of similar order of magnitude were 
 obtained, using different observables, by \cite{Hioki:2012vn,Englert:2012by,Biswal:2012dr,Baumgart:2012ay}.

 \subsection{Observables for tracing $\Rd$}
\label{suse:obscprd}

A non-zero chromo-electric  dipole moment  $\Rd$  induces CP-odd transverse
$\ttbar$ spin correlations
\cite{Bernreuther:1993hq,Bernreuther:1998qv},
 for instance $(\Sp\times\Sm)\cdot{\hk}_t$ (where ${\hk}_t$  is the
 top-quark direction of
 flight in the $\ttbar$ ZMF). 
These correlations generate, in the dileptonic decay modes
 the following CP-odd\footnote{As $|pp\rangle$ is not a CP eigenstate, a classification with
  respect to CP is, strictly speaking, not possible. However, as long as
  the acceptance cuts are CP-symmetric, the SM contributions to
  the expectation values $\langle {\cal O}_i\rangle$ are negligibly
  small. For a discussion, see \cite{Bernreuther:1998qv,Bernreuther:2010ny}.}  triple correlations \cite{Bernreuther:1998qv}:
\begin{equation}\label{O12}
{\cal O}_1 =  (\hlp \times \hlm) \cdot {\bf\hat k}_t \, , 
\qquad {\cal O}_2 = {\rm sign}(\cos\theta_t^*) \
(\hlp \times \hlm) \cdot  {\bf\hat p}  \, .
\end{equation}
The unit vectors $\hlp$, $\hlm$ 
 that refer to the   charged lepton directions of
 flight are defined as
above (cf. below \eqref{doublehel}),  while ${\bf\hat p}$
 is the direction of one of the proton beams (i.e., the $z$ axis)
 in the laboratory frame. The factor 
 ${\rm sign}(\cos\theta_t^*)$,  where $\cos\theta_t^* ={\bf\hat p}\cdot {\bf\hat
   k}_t$,  is the
 sign of the cosine of the top-quark scattering angle in
  the $\ttbar$ ZMF,  is required
  \cite{Bernreuther:1993hq,Bernreuther:1998qv} because the $gg$ initial state 
  is Bose symmetric.
 Without that factor, the second triple correlation in
 \eqref{O12} would have essentially no sensitivity to a non-zero
  $\Rd$.

The range of the correlations \eqref{O12} is
 $-1\leq {\cal O}_{1,2} \leq 1$. Within the SM, the distributions of
${\cal O}_{1,2}$ are symmetric around ${\cal O}_{1,2}=0$, i.e.,
 the expectation values $\langle {\cal O}_{1,2}\rangle_{SM}=0$ if no
acceptance cuts are applied or if these cuts are CP-symmetric.
 A non-zero $\Rd$ induces asymmetric distributions.

In the linear approximation, the expectation values of
 ${\cal O}_{1,2}$ are directly proportional to $\Rd$:
\begin{equation}\label{exO12}
\langle {\cal O}_{1,2}\rangle = c_{1,2} \ \Rd \, .
\end{equation}
Putting $\Rd =1$, these expectation values, i.e. the coefficients
 $c_{1,2}$ are given, for the LHC at 7 and 8 TeV,
 in Tables~\ref{tab:obsmw7cut} and~\ref{tab:obsmw8cut}, respectively,
 without and with a cut on $\mttbar$. In the computation of
 \eqref{exO12} and \eqref{asycp12} we have normalized to $\sigma_{LO}$.

Corresponding asymmetries are 
\begin{equation}\label{asycp12}
A^{CP}_i = \frac{ N_{\ell\ell}({\cal O}_i>0) -
   N_{\ell\ell}({\cal O}_i<0)}{N_{\ell\ell}} = \frac{9\pi}{16}
  \langle {\cal O}_i \rangle \, , \qquad i=1,2 \, .
\end{equation}
This relation between $A^{CP}_i$ and the corresponding expectation
value of ${\cal O}_i$ 
 holds if no acceptance cuts are applied, but is valid also if cuts on
 $\mttbar$ are made \cite{Bernreuther:1998qv}. Our predictions for these asymmetries are
  also collected in Tables~\ref{tab:obsmw7cut} and~\ref{tab:obsmw8cut}.

The highest sensitivity to  $\Rd$ would be obtained by fitting the
  distributions  $\sigma^{-1}d\sigma/d{\cal  O}_i$, which depend
  linearly on $\Rd$, to the respective unfolded experimental 
   distributions. One may expect to achieve a statistical uncertainty
    $\delta\Rd$ below the percent level (cf. Sect.~\ref{suse:Rmu}).
 For a crude estimate we use the above CP asymmetries. If
 $A^{CP}_1$ can be measured with 3 percent accuracy, then  $\Rd$
   may be determined  with an uncertainty 
  $\delta \Rd \simeq 0.04$.

 \begin{table} [h]
\begin{center}
\caption{Several expectation values of observables and asymmetries
 introduced in the text, without and with
  a cut on  $\mttbar$, for the LHC at 7 TeV. The chosen scale
 is $\mu=m_t$, the uncertainties in parentheses 
 are due to scale variations between $m_t/2$ and $2m_t$. The numbers
 are to be multiplied by the respective dimensionless chromo moment.}
\label{tab:obsmw7cut}
\vspace{2mm}
\begin{tabular}{lc|ccc}
\hline \hline 
  7 TeV &    &  $\mttbar \geq 2m_t$   &  $\mttbar \leq $ 450 GeV        &   $\mttbar>$ 450 GeV \\
\hline 
 $\langle {\cal O}_1\rangle$ &  $[\Rd]$ &
 $-0.397(10)$  & $-0.390(10)$  & $-0.403(10)$
\\
$A^{CP}_1$ & $[\Rd]$ & $-0.702(18)$ & $-0.689(18)$ & $-0.712(18)$   \\
$\langle {\cal O}_2\rangle$ & $[\Rd]$ & $-0.172(5)$ & $-0.104(2)$ & $-0.230(4)$
\\
$A^{CP}_2$ & $[\Rd]$ & $-0.304(9)$ & $-0.184(4)$  & $-0.406(7)$   \\
 $\langle {\cal O}_T\rangle_{NP}$ & $[\Imu]$ & $0.057(4)$ &  $0.088(4)$
& $0.031(5)$ \\
$A_T^{NP}$  & $[\Imu]$ &  $0.114(8)$ &  $0.176(8)$ & $0.062(10)$  \\
$\langle {\cal O}_T\rangle_{QCD}$ &  & $0.0026(6)$ & $0.0012(4)$ & $0.0038(4)$
\\
$A_T^{QCD}$ & & $0.0052(12)$ & $0.0024(8)$ & $0.0076(8)$ \\
\hline \hline
\end{tabular}
\end{center}
\end{table}

 \begin{table} [h]
\begin{center}
\caption{Same as Table~\ref{tab:obsmw7cut}, for the LHC at 8 TeV.}
\label{tab:obsmw8cut}
\vspace{2mm}
\begin{tabular}{lc|ccc}
\hline \hline 
  8 TeV &    &  $\mttbar \geq 2m_t$   &  $\mttbar \leq $ 450 GeV        &   $\mttbar>$ 450 GeV \\
\hline 
 $\langle {\cal O}_1\rangle$ &  $[\Rd]$ &
 $-0.415(10)$  & $-0.407(10)$  & $-0.420(10)$
\\
$A^{CP}_1$ & $[\Rd]$ & $-0.734(18)$ & $-0.720(17)$ & $-0.743(17)$   \\
$\langle {\cal O}_2\rangle$ & $[\Rd]$ & $-0.180(4)$ & $-0.107(2)$ & $-0.237(4)$
\\
$A^{CP}_2$ & $[\Rd]$ & $-0.318(7)$ & $-0.189(3)$  & $-0.419(7)$   \\
$\langle {\cal O}_T\rangle_{NP}$ & $[\Imu]$ & $0.068(4)$ &  $0.010(2)$
& $0.047(4)$ \\
$A_T^{NP}$  & $[\Imu]$ &  $0.136(8)$ &  $0.020(4)$ & $0.094(8)$  \\
$\langle {\cal O}_T\rangle_{QCD}$ &  & $0.0026(3)$ & $0.0012(2)$ & $0.0038(2)$
\\
$A_T^{QCD}$ & & $0.0052(6)$ & $0.0024(4)$ & $0.0076(4) $ \\
\hline \hline
\end{tabular}
\end{center}
\end{table}

 \subsection{Longitudinal polarization and  $\Id$}
\label{suse:lopol}

If the chromo-electric dipole moment of the top quark has a non-zero
imaginary part, $\Id \neq 0$, then 
 $P$- and CP-odd contributions (that are $T_N$-even)  to
the  $q{\bar q}$- and $gg$-initiated matrix elements are induced. This leads to
 a longitudinal polarization of both the $t$ and $\bar t$ in the
  $\ttbar$ sample \cite{Bernreuther:1993hq,Bernreuther:1998qv},
for instance with respect to the $t$ and $\bar t$ directions of flight
in the $\ttbar$ ZMF: These CP-odd contributions lead to
$\langle\Sp\cdot{\bf\hat{k}_t}\rangle = 
\langle\Sm\cdot{\bf\hat{k}_{\bar t}}\rangle \propto 
    \Id$.  The P-violating SM interactions\footnote{As in the case of
      \eqref{O12}, the contribution of
 the Kobayashi-Maskawa  phase is completely negligible.} and possibly
new P-violating, but CP-conserving interactions also lead to
 a longitudinal $t$ and $\bar t$ polarization -- in this case one gets
 $\langle\Sp\cdot{\bf\hat{k}_t}\rangle = -
\langle\Sm\cdot{\bf\hat{k}_{\bar t}}\rangle.$

A search for an non-zero longitudinal  $t$ and $\bar t$ polarization
is most efficiently made in the lepton + jets decay channels, because
i) only one charged lepton  is required as analyzer of
the $t$ and $\bar t$ spin, respectively, and ii) the $t$ and $\bar t$
rest frames can be reconstructed quite efficiently for these channels. 
If one uses the lepton helicity angles introduced below
\eqref{doublehel},
 i.e.,  $\theta_1 =\angle(\hlp,{\bf\hat k}_t)$ and
 $\theta_2 =\angle(\hlm,{\bf\hat k}_{\bar t})$, then
 one obtains, for the reactions  \eqref{ppljt},  \eqref{ppljtbar} 
    the distributions (if 
 no acceptance cuts are applied)
\begin{equation}\label{discpdim}
\sigma^{-1}\frac{d\sigma}{d\cos\theta_{1,2}} =
 \frac{1}{2}\left(1 + B_{1,2}\cos\theta_{1,2}\right) \, ,
\end{equation}
where
\begin{equation}  \label{disBdim}
     B_{1,2} = B_{SM} \pm B_{NP} \Id  \, .
\end{equation}
The longitudinal $t$ and $\bar t$ polarization induced by SM
parity-violating weak interactions is very small; we collect it for
reference in Table~\ref{table:Bhel}, where
  our results for  $B_{NP}$ are also listed. (For the computation of
 $B_{SM}$ we took into account our results
 \cite{Bernreuther:2006vg,Bernreuther:2008md}.)
In the computation of
   $B_{1,2}$  we have normalized to $\sigma_{LO}$.
\begin{table}[htb]
\caption{The SM and NP contributions
   to the longitudinal $t$ and $\bar t$ polarization  \eqref{disBdim},
   respectively, for $\mu=m_t$.  The uncertainties in parentheses 
 are due to scale variations between $m_t/2$ and $2m_t$.}
\label{table:Bhel}
\begin{center}
\begin{tabular}{ll|ccc}
\hline \hline 
  &  & $\mttbar\geq 2m_t$ & $\mttbar \leq 450$ GeV    & $\mttbar > 450$ GeV \\
\hline 
7 TeV:& $B_{SM}$ &   $0.003(1)$ & $0.001(1)$ & $0.005(1)$ \\
    & $B_{NP}$ &  $0.497(4)$ & $0.497(5)$ &$0.498(6)$ \\
 & ${\tilde B}_{NP}$ & $0.409(12)$ & $0.493(11)$ & $0.339(12)$ \\
\hline
8 TeV: & $B_{SM}$  & $0.003(1)$ & $0.001(1)$ & $0.005(1)$ \\
 & $B_{NP}$ &  $0.482(3)$  & $0.491(4)$    & $0.474(5)$ \\
 & ${\tilde B}_{NP}$ & $0.421(10)$ & $0.510(10)$ & $0.351(12)$ \\
\hline \hline 
\end{tabular}
\end{center}
\end{table} 
Acceptance cuts on the transverse momentum and on the rapidity of the
 charged leptons severely distort the  shape of the distributions 
\eqref{discpdim} in the backward region $\cos\theta_{1,2}\leq 0$
(cf. e.g. \cite{Bernreuther:2010ny}.) In order to probe for a non-zero $\Id$ it may
therefore be appropriate in an experimental analysis to
 measure the distributions \eqref{discpdim}, both for the $\ell^+$ and
 $\ell^-$ data samples, only in the forward region and consider the
 ``asymmetry'' 
\begin{equation}\label{Asyimd}
 A_P = \frac{N_{\ell^+}(\cos\theta_1>0)}{N_{\ell^+}} -
  \frac{N_{\ell^-}(\cos\theta_2>0)}{N_{\ell^-}} =  \frac{1}{2} B_{NP} \Id  \, .
\end{equation}

Alternatively, one may  search for a longitudinal $t$ and $\bar t$
polarization  with the respect to the direction $\hp$ of one of the
proton beams in the lab frame. In this case $\Id\neq 0$ leads to 
\begin{equation}\label{lopolbea}
\langle    {\rm sign}(\cos\theta_t^*)      \Sp\cdot\hp\rangle = 
- \langle{\rm sign}(\cos\theta_t^*)\Sm\cdot\hp\rangle \propto 
    \Id \, , 
\end{equation}
where the factor ${\rm sign}(\cos\theta_t^*)$ is required because of
the Bose symmetry of the  gluon-fusion matrix element  squared, cf.
Sect.~\ref{suse:obscprd}. This polarization leads, in the
  $\ell^\pm$ + jets final states, to non-flat distributions of 
\begin{equation}\label{lopolbea2} 
\cos{\tilde\theta}_1 = \hp\cdot\hlp \, , \qquad \cos{\tilde\theta}_2
= - \hp\cdot\hlm \, , 
\end{equation}
(notice the minus sign in the definition of  $\cos{\tilde\theta}_2$), in analogy to \eqref{discpdim}.
In analogy to \eqref{Asyimd} one may consider 
\begin{equation}\label{Asyimd2}
{\tilde A}_P  = \frac{N_{\ell^+}(\cos{\tilde\theta}_1>0)}{N_{\ell^+}} -
  \frac{N_{\ell^-}(\cos{\tilde\theta}_2>0)}{N_{\ell^-}} = \frac{1}{2} {\tilde B}_{NP} \Id \, ,
\end{equation}
for  $\ell^\pm$ + jets final states, separately
 for events  with ${\rm sign}(\cos\theta_t^*)=1$ and $-1$.   The
 coefficients  ${\tilde B}_{NP}$ are also collected in Table~\ref{table:Bhel}.

If no acceptance cuts are made, our results for the distributions
\eqref{discpdim}
 and the analogous ones for $\cos{\tilde\theta}_{1,2}$ apply also to
 dileptonic final states.

The longitudinal top polarization was first measured by the
D$\emptyset$ collaboration \cite{Abazov:2012oxa} at the Tevatron.
 At the LHC (7 TeV) the longitudinal top polarization  $p_t$  was measured
in the helicity basis by CMS \cite{CMS:2012owa} for dilepton events and by ATLAS
\cite{ATLAS:2012epa} for $\ell$ + jets events.  These measurements are
compatible with zero and the achieved precision (statistical and
systematic uncertainties added  in quadrature) is $\delta p_t \simeq
0.05$. Using that the top-spin analyzing power of $\ell^+$ is
$\kappa_{\ell}=1$ to very good approximation \cite{Brandenburg:2002xr}, the polarization $p_t$
 as defined in  \cite{CMS:2012owa} is related to \eqref{discpdim}
 by $p_t=B_1/2$. From  $\delta p_t^{exp} \simeq
0.05$ and \eqref{discpdim} one obtains, roughly, the upper bound
$|\Id|\lesssim 0.2$. This bound may be improved by exploration of the 8
TeV data. A sensitivity estimate of similar order of magnitude was
obtained by \cite{Biswal:2012dr}.

 \subsection{Transverse polarization and  $\Imu$}
\label{suse:trapol}

 The  SM-induced longitudinal polarization of the $t$ and
 $\bar t$ quarks of the hadronically produced $\ttbar$ sample is very
 small, and also their polarization transverse to the scattering plane
 due to QCD absorptive parts of the $q{\bar q}$- and
 $gg$-induced scattering amplitudes is below 1 percent. A complete calculation to order
 $\alpha_s^3$ of this polarization was made in \cite{Bernreuther:1995cx,Dharmaratna:1996xd}. 
 In addition, this P- and CP-even, $T_N$-odd polarization receives also
 a contribution from a non-zero imaginary part of the CMDM of the
 top quark,  $\Imu\neq 0$.

Again, the lepton +
 jets final states are obviously  the most suitable ones to search for this effect.
 Appropriate observables are constructed as follows. 
 We define a vector  ${\bf n}$ that is perpendicular to the
 $ij\to \ttbar$ scattering plane,  ${\bf n} ={\bf\hat p}\times {\bf\hat
   k}_t$, and consider, for $\ell^+$ $(\ell^-)$ + jets events, the
 correlation of the  $\ell^+$ $(\ell^-)$ direction of flight in the
 $t$ $({\bar t})$ rest frame with  ${\bf n}.$ This is achieved, for
 the respective final states, with the observables
 \begin{equation}\label{ObsTabs}
 {\cal O}_T =  {\rm sign}(\cos\theta_t^*) \ {\bf n}\cdot \hlp \, ,  \qquad 
  {\overline{\cal O}}_T =  - {\rm sign}(\cos\theta_t^*) \ {\bf n}\cdot
  \hlm  \, .
\end{equation}
The factor  ${\rm sign}(\cos\theta_t^*)$ is necessary for
 obtaining non-zero expectation values of ${\cal O}_T$ and
 ${\overline{\cal O}}_T$, cf. Sect.~\ref{suse:obscprd} 
 and\footnote{Our observables \eqref{ObsTabs} 
  have a significantly higher sensitivity than the corresponding
 lab-frame observables used in  \cite{Bernreuther:1995cx}.} \cite{Bernreuther:1995cx}.
 We have 
\begin{equation}\label{MWabsT}
\langle{\overline{\cal O}}_T \rangle = \langle{{\cal O}}_T \rangle=
 \langle{{\cal O}}_T \rangle_{QCD} + \langle{{\cal O}}_T\rangle_{NP} \, ,
\end{equation}
and in the linear approximation for the chromo moments, the NP
contribution is directly proportional to  $\Imu$:
 $\langle{{\cal O}}_T\rangle_{NP} = c_T \Imu .$
In addition one may consider the asymmetry
\begin{equation}\label{Asyabs}
 A_T = \frac{N_\ell({\cal
    O}_T>0)-N_\ell({\cal O}_T<0)}{N_\ell} = 2~\langle{\cal O}_T \rangle  \, ,
\end{equation}
and likewise for ${\overline{\cal O}}_T$,  where the equality on the right-hand side holds in the absence of
acceptance cuts.
Our predictions of the contributions from QCD and from
 a non-zero $\Imu$ to the expectation value
 \eqref{MWabsT}      and to the   asymmetry    \eqref{Asyabs} are
  collected in Tables~\ref{tab:obsmw7cut} and~\ref{tab:obsmw8cut},
 both without and with a cut on the $\ttbar$ invariant mass.
In the computation of  \eqref{MWabsT}    and \eqref{Asyabs}  we have normalized to $\sigma_{LO}$.

Plenty of $\ttbar$ lepton + jets events were recorded at the LHC.
 For instance, ATLAS has selected  close to
 $4\times 10^4$ such events $(\ell=e,\mu)$ from their 7 TeV
 data. Thus, one expects $\sim 16 \times 10^4$ such events  from
 the 8 TeV (20 fb$^{-1}$) data. The ratio of events with $\mttbar\leq 450$
 GeV and $\mttbar>450$ GeV is approximately $0.45 : 0.55$. Thus, the
 statistical uncertainty  in measuring the asymmetry \eqref{Asyabs} is
 expected to be below $1\%$. If a combined uncertainty 
 $\delta A_T^{exp} \simeq 0.02$ can be achieved, both for low and high $\mttbar$ events,
  $\Imu$ may be determined with an uncertainty of $\delta\Imu\simeq
  0.12$. A higher precision can be obtained by exploiting the
  distributions of \eqref{ObsTabs}.

\section{Summary}
\label{sec:concl}

Top-spin observables are becoming a useful tool for the detailed
exploration of $\ttbar$ production and decay at the LHC. In view of
the large $\ttbar$ data samples that were recorded by the ATLAS and
CMS collaborations at LHC center-of-mass energies of 7 and 8 TeV and that
are presently being analyzed  in detail, we have extended our
previous NLO SM predictions of a number of top-spin correlation and
polarization observables to these energies. The transverse polarization of $t$ and $\bar t$
quarks is an interesting probe of the $\ttbar$
 production dynamics -- the QCD induced asymmetry   $A_T^{QCD}$ is, however, only
 $0.76 \%$ at 7 TeV and 8 TeV for events with $\mttbar>$ 450 GeV.
 
In addition, we have parameterized possible new physics effects
 in the hadronic $\ttbar$ production matrix elements by complex
 chromo-magnetic and chromo-electric dipole moments of the top quark.
 Using the presently available empirical information about these
 moments, namely, that the moduli of the respective dimensionless
 moments  must be markedly smaller than one, we have analyzed
 a number of charged lepton angular correlations and distributions
 that allow to disentangle the effects of a non-zero
 $\Rmu, \Rd, \Imu,$ and $\Id$.  We expect that the analysis of
  the available  7 and 8 TeV $\ttbar$ data samples by ATLAS and CMS
  will allow to achieve, at least for the real parts
 of these dimensionless moments,  a  sensitivity of a few
 percent.  This would provide significant
  direct information on whether or not hadronic top-quark production
 is affected by new  interactions at length scales as small as
 $10^{-18}$ cm.


\subsubsection*{Acknowledgements}

 We wish to thank J. Andrea, F. Deliot, D. Heisler, F. H\"ohle,
 C. Schwanenberger, and E. Yazgan for
  discussions.  The work of  W.B.  was supported by BMBF and that of Z.G. Si  by NSFC and by Natural Science Foundation of
Shandong Province.


\end{document}